\documentclass[pre,twocolumn,showpacs]{revtex4}

\usepackage{graphicx}

\usepackage{amsmath,amssymb}
\usepackage{bm}
\usepackage{cancel}

\begin{document}
\title{Hybrid Modeling of Tumor-induced Angiogenesis}
\author{ L.\ L.\ Bonilla$^1$, V.\ Capasso$^{1,2}$, M.\ Alvaro$^1$ and M.\ Carretero$^1$}
\affiliation {$^1$G. Mill\'an Institute, Fluid Dynamics, Nanoscience
and Industrial Mathematics, Universidad Carlos III de Madrid, 28911
Legan\'es, Spain} \affiliation{$^2$ ADAMSS, Universit\'a degli Studi
di Milano, 20133 MILANO, Italy}

\date{\today}
\begin{abstract}
When modeling of tumor-driven angiogenesis, a major source of analytical and computational complexity is the strong coupling between the kinetic parameters of the relevant stochastic branching-and-growth of the capillary network, and the family of interacting underlying fields. To reduce this complexity, we take advantage of the system intrinsic multiscale structure: we describe the stochastic dynamics of the cells at the vessel tip at their natural mesoscale, whereas we describe the deterministic dynamics of the underlying fields at a larger macroscale. Here, we set up a conceptual stochastic model including branching, elongation, and anastomosis of vessels and derive a mean field
approximation for their densities. This leads to  a deterministic integro-partial differential system that describes the formation of the stochastic vessel network. We discuss the proper capillary injecting boundary conditions and include the results of relevant numerical simulations.
\end{abstract}

\maketitle

\section{Introduction}

\begin{figure}[ht]
\begin{center}
\includegraphics[height=8cm]{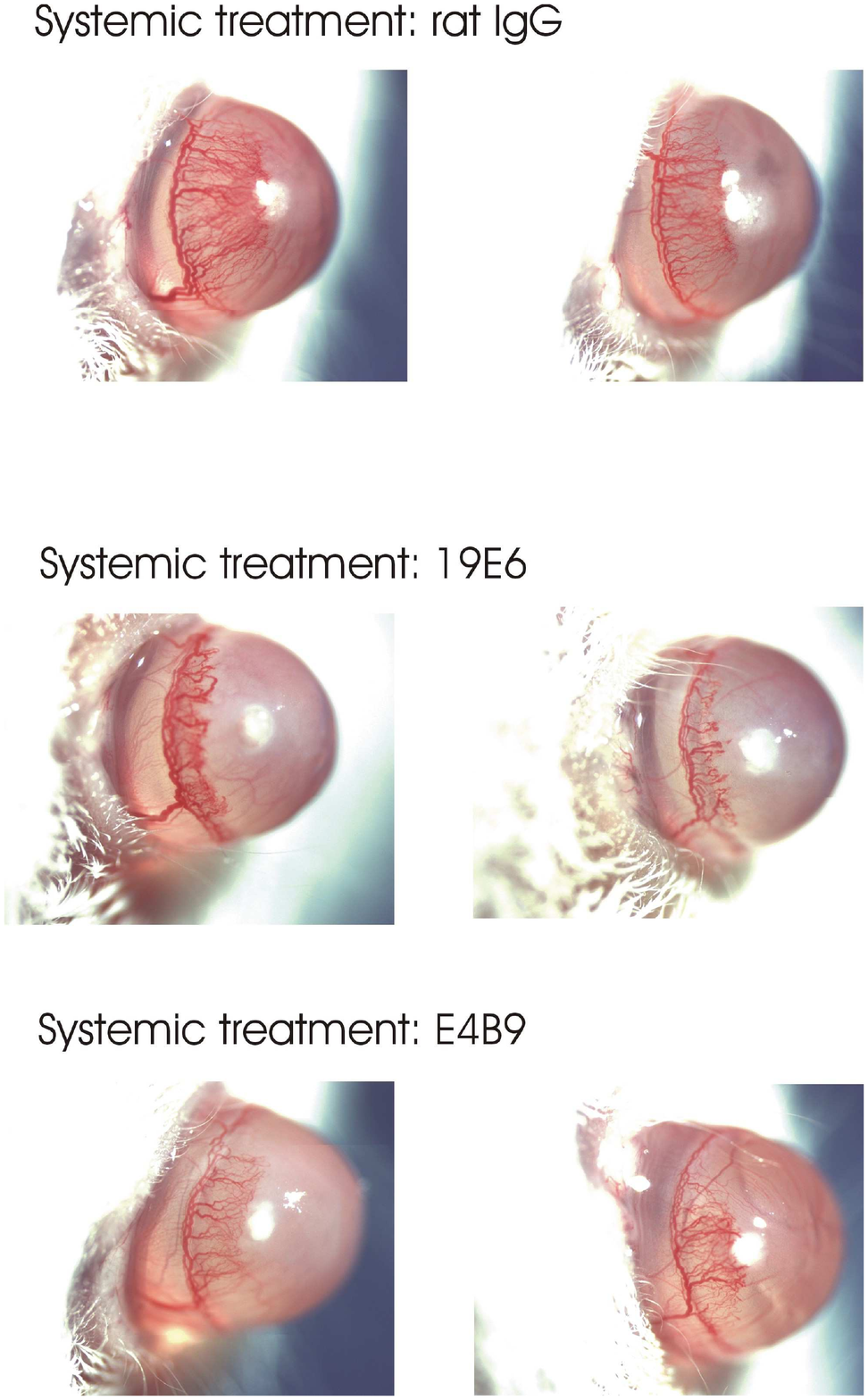} \qquad
\end{center}
\caption{Angiogenesis on a rat cornea. The photographs show the angiogenic response to a cornea injury after different anti-angiogenic treatments that inhibit vessel extension and proliferation. Photographs courtesy of E. Dejana.
\label{angio_real2}}
\end{figure}

The growth of blood vessels (a process known as angiogenesis) is
essential for organ growth and repair. An imbalance in this process
contributes to numerous malignant, inflammatory, ischaemic,
infectious and immune disorders; according to Carmeliet
\cite{carmeliet_2005} ``angiogenesis research will probably change
the face of medicine in the next decades, with more than 500 million
people worldwide predicted to benefit from pro- or anti-
angiogenesis treatments". In particular, while angiogenesis does not
initiate malignancy, it promotes tumor progression and metastasis \cite{GG2005,CT2005,CJ2011}.
Viceversa a large effort has been recently  dedicated to analyzing
the effects of anti-angiogenic therapies to reduce, and possibly
eliminate, tumor growth. In this context a quantitative approach is
crucial, since  therapy can be interpreted mathematically as an
optimal control  problem, where  the effort  of the anti-angiogenic
treatment has to  be confronted with its costs, and its
effectiveness. Experimental dose/effect analysis  are nowadays
routine in many biomedical laboratories (see  e.g.
\cite{morale:folkman_1974,jain-carmeliet,corada-dejana} and Figure
\ref{angio_real2}), but still they lack methods of optimal
control, which are typical of engineering and economic systems. An
interesting numerical investigation has been carried out in
\cite{harrington} regarding a model of  tumor induced angiogenesis \cite{tong}
subject to inhibitors. On the other hand methods of optimal control
require a solid underlying mathematical model which has to be
validated by real experiments (see  e.g. \cite{burgerVKAM}).

 An important contribution has come from the experiments and related quantitative analysis reported in \cite{morale:Stokes_Lauffenburger:91bis,morale:Stokes_Lauffenburger:91}, where the authors emphasize the importance of a ``probabilistic framework, capable of simulating the development of individual microvessels and resulting networks". Actually a   angiogenic system is extremely complex  due to its intrinsic multiscale structure. When modeling such systems, we need to consider the strong coupling between the kinetic parameters of the relevant microscale branching and growth stochastic processes of the capillary network and the family of interacting macroscale underlying fields. Capturing the keys of the whole process is still an open problem while there are many models in the literature that address some partial features of the angiogenic process \cite{morale_chaplain_1998,morale:plank_sleeman:03,othmer,morale:sun_wheeler:05,morale:sun_wheeler_SIAM:05,morale:chaplian_2006,VK_morale_jomb,swanson2011,preziosi,cotter}. 
 
 Hybrid models reduce complexity exploiting the natural multiple scale nature of the angiogenic system. Often hybrid models treat vessel cells on the extracellular matrix as discrete objects, and different cell processes like migration, proliferation, etc. occur with certain probabilities. The latter depend on concentrations of certain chemical factors; these concentrations satisfy reaction-diffusion equations (RDEs) \cite{morale_chaplain_1998,harrington,tong,preziosi}. In other approaches, the cell microscale is not treated explicitly. In a mesoscale, large compared to cell size but small compared to the macroscale of the concentrations, vessels are wires that move and grow randomly toward the tumor by chemotaxis \cite{VK_morale_jomb,MCO2}. An important simplifying factor is that the stalk cells in a growing vessel build the  capillary following the wake of the cells at the vessel tip \cite{CJ2011}. Thus the idealized wire that follows a vessel tip may be assumed to comprise all previous positions of the vessel tip. In this way only the simple stochasticity of the geometric processes of birth (branching) and growth is kept. We can then focus our attention on the random evolution of tip vessels and their coupling with the underlying concentration fields that interact with them \cite{VK_morale_jomb,MCO2}. 
 
The RDEs for the underlying fields contain terms that depend on the spatial distribution of vascular cells. Our idea is use a mean field  approximation for cell distribution so that, in the limit of large number of cells, the underlying fields become deterministic. The full multiscale mesoscopic model of angiogenesis consists of a stochastic description of vessel tips coupled to RDEs containing mean field terms that depend on the distribution of vessels. The latter are random and therefore the equations for the underlying fields are stochastic. A {\em hybrid model} consists of approximating the random RDEs by deterministic ones in which the terms depending on cell distributions are replaced by their averages. Once the governing equations of the model are established, its parameters can be estimated from data and their effect on the solution of the model ascertained. This could help assessing anti-angiogenic therapies that control vascularization. Figure \ref{angio_real2} shows the angiogenic response to injuries in a rat cornea in the presence of different drugs. If one can correlate the effect of the drugs on the parameters of the hybrid model or identify drug presence with some additional terms, optimal control of the equations may help devising the most appropriate therapies. 

The  importance of using an intrinsically  stochastic model at  the microscale to describe the generation of  a  realistic vessel network has been the subject of  a series  of papers by one of the present authors \cite{VK_2014,VK_morale_facc_2013}. Complementary to  the direct problem of
modeling an angiogenic network, the statistical problem of estimating spatial densities of fybers in a random network, has been faced in \cite{indam,cam_VK_villa}. The statistical problem has great
importance for validating the direct models on the basis of images taken from experiments, such as those shown in Figure \ref{angio_real2}.

Here we are emphasizing the problems related to the mean field description of the underlying biochemical fields. In the literature there are examples of rigorous derivations of mean field equations  from stochastic particle dynamics \cite{oel2,sznitman,C_M,B_C_M}. However, to the best of the authors' knowledge, the kind of stochastic hybrid models considered here have not yet been
studied and require further investigation.

Here  we derive the above mentioned  mean field approximation from a conceptual stochastic model   for the formation of the stochastic network of vessels. Using heuristic  arguments, we show that the
spatial distribution of  the tip  density satisfies a nonlinear integrodifferential  evolution equation coupled  with the partial differential equations for the relevant  underlying fields.

We  start from an extension of the mathematical model proposed in \cite{VK_morale_jomb}, according to which (see e.g. \cite{morale:folkman_1974,morale:Stokes_Lauffenburger:91,morale:Stokes_Lauffenburger:91bis,morale:plank_sleeman:03,morale:sun_wheeler_SIAM:05}) the endothelial cells proliferate and migrate in response to different signaling cues. The motion of endothelial cells is led by cells at the vessel tip, whereas other cells follow doggedly the tips and form the vessel. Thus we can track the motion of the vessel tips and the vessels are simply the trajectories thereof. Vessel tips move along gradients of a diffusible substance and a growth factor emitted by the tumor (tumor angiogenetic factor, TAF). Thus their motion is controlled by chemotaxis and, in addition, by haptotaxis, the directed cell movement along an adhesive gradient (here fibronectin) of a non diffusible substance. Specific biochemical mechanisms are  widely described in literature (see e.g. \cite{morale:Stokes_Lauffenburger:91bis}).

Two additional mechanisms are responsible for the formation of the vessel network: tip branching (here assumed to occur only at existing tips for the sake of simplicity) and anastomosis that occurs whenever a tip runs into another existing vessel, merges with it and stops moving. Both mechanisms are intrinsically random. Tip branching is a birth process driven by the underlying fields mentioned above. In this paper, we have included a model of anastomosis as a death process of a tip that encounters an existing vessel and is therefore coupled with the density of the vessel network. This is a significant improvement with respect to the previous work \cite{VK_morale_jomb}.

We have derived formally the mean field equation for the spatial  density of tips, which is  a  function of  tip  location and velocity. This equation is a parabolic integrodifferential equation of Fokker-Planck type having a source term and a noninvertible diffusion matrix: it is second order in the derivatives  with respect to the velocities and first order in the derivatives with respect to the position coordinates. Together with the mean field equations for the underlying fields, we have thus found an independent integrodifferential system whose solution will provide the required (now deterministic) parameters which drive the stochastic system for the tips, eventually leading to the stochastic vessel network, at the microscale. These  arguments confirm the need by itself of an accurate analysis of the mean field approximation of the underlying fields. 

The main scope of this paper is to establish an  adequate initial-boundary value problem (IBVP) for the integrodifferential system. Due to the peculiar structure thereof, the choice of boundary  conditions is crucial. In this paper, we introduce novel boundary conditions based on the physical situation we model and also on related ideas used to describe the injection of electric charge through contacts of semiconductor devices \cite{BGr05,CBC09,alvaro13}. We do not study here whether the IBVP is well-posed; see \cite{ana}. Instead, we have explored its qualitative behavior by numerically solving the IBVP for TAF concentration and tip density. These numerical solutions confirm what is expected from the model.

The rest of the paper is as follows. Section \ref{sec:model} describes how our stochastic model treats vessel branching, extension and anastomosis. We derive the equation of Fokker-Planck type for the density of vessel tips and the TAF RDE in Section \ref{morale:sec_evolution}. The appropriate boundary and initial conditions are proposed and discussed in Section \ref{sec:bc}. Numerical results for the nondimensional version of the equations are reported in Section \ref{sec:numerical} whereas section \ref{sec:conclusions} contains our conclusions. Appendix \ref{ap1} is devoted to mathematical details that are used to derive the Fokker-Planck type equation.

\section{The mathematical model} \label{sec:model}
Based on the above discussion, the main features of the process of
formation of a  tumor-driven  vessel network are (see
\cite{morale:chaplain_stuart:93,morale:plank_sleeman:03,VK_morale_jomb})
\begin{itemize}
\item[i)] vessel branching;

\item[ii)] vessel extension;

\item[iii)] chemotaxis in response to a generic tumor
angiogenetic factor (TAF), released by   tumor cells;

\item[iv)] haptotatic migration in response to fibronectin
gradient, emerging from the extracellular matrix and through
degradation and production by endothelial cells themselves;

 \item[v)] anastomosis, when a capillary tip meets an existing vessel.

\end{itemize}

\medskip

\noindent Let   $N_0$ denote the initial number of tips, $N(t)$ the numbers of tips at time $t$, $\mathbf{X}^i(t)$ the location of the $i$-th tip at time $t$, and $\mathbf{v}^k(t)$ its velocity. We model sprout
extension by tracking the trajectory of individual capillary tips.

 \subsubsection{Tip branching}
We assume that vessels branch out of moving tips and ignore branching from mature vessels. A tip $i$ is born at a random time $T^i$ and disappears at a later random time $\Theta^i$, either by reaching the tumor or by anastomosis (see later). We assume that the probability that a tip branches from one of the existing ones during an infinitesimal time interval $(t, t + dt]$ is 
\begin{eqnarray}
\sum_{i=1}^{N(t)}\alpha(C(t,\mathbf{X}^i(t)))\, dt,  
\label{prob_nuova_nascita_tip}
\end{eqnarray}
where $\alpha(C)$ is a non-negative function of the TAFÕs concentration $C(t,\mathbf{x})$. For example, we  may  take
\begin{eqnarray}
\alpha(C)=\alpha_1\frac{C}{C_R+C},
\end{eqnarray}
where $C_R$ is a reference density parameter \cite{VK_morale_jomb}. The evolution equation for $C(t,\mathbf{x})$ will be given later. As a technical simplification, we will further assume that whenever a tip located in $\mathbf{x}$ branches, the initial value of the state of the new tip is $(\mathbf{X}^{N(t)+1},\mathbf{v}^{N(t)+1}) = (\mathbf{x},\mathbf{v}_0)$, where $\mathbf{v}_0$ is a non random velocity. 

\subsubsection{Vessel extension}
Vessel extension is described by the Langevin equations
\begin{eqnarray} d\mathbf{X}^k(t)&=&\mathbf{v}^k(t)\, dt\nonumber\\
d\mathbf{v}^k(t)&=& \left[- k\, \mathbf{v}^k(t)+
\mathbf{F}\!\left(C(t,\mathbf{X}^k(t))\right)\!
\right]\! dt \nonumber \\
&+& \sigma\, d\mathbf{W}^k(t), \label{eq:langevin}
\end{eqnarray}
(for $t>T^k,$ the random  time  at which the $k$th tip appears). Besides the friction force, there is a force due to the underlying TAF field $C(t,\mathbf{x})$ \cite{morale:plank_sleeman:03,morale:chaplian_2006}: 
\begin{eqnarray}
\mathbf{F}(C)= \frac{d_1}{(1+\gamma_1C)^q}\nabla_x C.
\label{Fvector}
\end{eqnarray}
We are ignoring other processes such as production and degradation of other fields such as fibronectin and matrix degrading enzyme (MDE) that further complicate the model. 

\subsubsection{Anastomosis}
When a vessel tip meets an existing vessel it joins it at that point and time and it stops moving. This {\em death} process is called tip-vessel {\em anastomosis}.

\section{The evolution of the  empirical measures associated with the  tip process.}
\label{morale:sec_evolution}
Let us now derive the governing equations of the model. We shall first ignore branching and consider only vessel extension given by (\ref{eq:langevin}). Later we will consider the effects of tip branching and anastomosis. 

\paragraph{Vessel extension.} Let $g(x,v)$ be a smooth test function. By Ito's formula (see p.\! 93 of \cite{gardiner10} or p.\! 252 of \cite{ capasso_bakstein}), we get from (\ref{eq:langevin}),
\begin{eqnarray}
&&dg(\mathbf{X}^k(s),\mathbf{v}^k(s))=\! \mathbf{v}^k(s)\cdot\nabla_{x}g(\mathbf{X}^k(s),\mathbf{v}^k(s))ds \nonumber\\
&&+[\mathbf{F}(C(s, \mathbf{X}^k(s)))-k\mathbf{v}^k(s)]\cdot\nabla_v g(\mathbf{X}^k(s),\mathbf{v}^k(s))ds \nonumber \\
&&+\frac{\sigma^2}{2}\Delta_v g(\mathbf{X}^k(s),\mathbf{v}^k(s)) ds\nonumber\\
&&+ \nabla_v g(\mathbf{X}^k(s),\mathbf{v}^k(s))\cdot
d\mathbf{W}^k(s). \label{gi}
\end{eqnarray}
We now assume that $N$ is a fixed positive parameter of the same order as the number of tips $N(t)$ that may be counted during an experiment. Using now
$$g(\mathbf{X}^k(t),\mathbf{v}^k(t))\!=\!g(\mathbf{X}^k(0),\mathbf{v}^k(0))\!+\!\int_0^t dg(\mathbf{X}^k(s),\mathbf{v}^k(s)),$$ 
we deduce
\begin{eqnarray}
&& \frac{1}{N} \sum_{k=1}^{N(t)}g(\mathbf{X}^k(t),\mathbf{v}^k(t))=\frac{1}{N} \sum_{k=1}^{N(t)}g(\mathbf{X}^k(0),\mathbf{v}^k(0))\nonumber \\
&&
+ \int_0^t\!\frac{1}{N} \sum_{k=1}^{N(s)}\!\mathbf{v}^k(s)
\cdot\nabla_xg(\mathbf{X}^k(s),\mathbf{v}^k(s))\,ds \nonumber \\
 &&+ \int_0^t\!\frac{1}{N} \sum_{k=1}^{N(s)} [\mathbf{F}(C(\mathbf{X}^k(s)))-k\mathbf{v}^k(s)] \nonumber \\
  && \quad \quad \cdot\nabla_v g(\mathbf{X}^k(s),\mathbf{v}^k(s))\,ds
 \nonumber\\
&&\!+\frac{\sigma^2}{2}\! \int_0^t\!\!\frac{1}{N} \sum_{k=1}^{N(s)}
\Delta_v g(\mathbf{X}^k(s),\mathbf{v}^k(s))\, ds +
\tilde{M}_{1,N}(t),\label{gN}\end{eqnarray}
where
\begin{eqnarray}
\tilde{M}_{1,N}(t)= \!\int_0^t\!\frac{1}{N}
\sum_{k=1}^{N(s)}\!\nabla_v g(\mathbf{X}^k(s),\mathbf{v}^k(s))\cdot d\mathbf{W}^k(s),\label{mtilde}
\end{eqnarray}
is a zero mean martingale with $\tilde{M}_{1,N}(t)\to 0$ as $N\to\infty$; see p.\ 185 of \cite{capasso_bakstein}. In the limit as $N\to\infty$, we may write
\begin{equation}
 \frac{1}{N} \sum_{k=1}^{N(t)}g(\mathbf{X}^k(t),\mathbf{v}^k(t))\sim \int g(\mathbf{x},\mathbf{v})\, p(t,\mathbf{x},\mathbf{v})\, d\mathbf{x}d\mathbf{v},\label{p_def}
\end{equation}
where $p(t,\mathbf{x},\mathbf{v})$ is the tip density at time $t$. Then Eq.\!\! (\ref{gN}) can be written as
\begin{eqnarray}
&&\int\! g(\mathbf{x},\mathbf{v})\,
p(t,\mathbf{x},\mathbf{v})\, d\mathbf{x}d\mathbf{v}\!
=\!\int\! g(\mathbf{x},\mathbf{v})\, p(0,\mathbf{x},\mathbf{v})\, d\mathbf{x}d\mathbf{v} \nonumber \\
&&+ \int_0^t\!\!\!\int\!\!p(s,\mathbf{x},\mathbf{v})\mathbf{v}\cdot\nabla_xg(\mathbf{x},\mathbf{v}) d\mathbf{x}d\mathbf{v}\, ds \nonumber\\
&&+\!
\int_0^t\!\!\!\int\!p(s,\mathbf{x},\mathbf{v})[\mathbf{F}(C(s,\mathbf{x}))\!-\!k\mathbf{v}]\!\cdot\!\nabla_v
g(\mathbf{x},\mathbf{v}) d\mathbf{x}d\mathbf{v} ds
\nonumber \\
&&+ \int_0^t\!\! \!\int\!\frac{\sigma^2}{2}p(s,\mathbf{x},\mathbf{v})\Delta_v g(\mathbf{x},\mathbf{v})\, d\mathbf{x}d\mathbf{v}ds.\label{g}
\end{eqnarray}
Integrating by parts this equation and time differentiating the result, we obtain the Fokker-Planck equation for $p$:
\begin{eqnarray}
\frac{\partial}{\partial t}p(t,\mathbf{x},\mathbf{v})&=&-\mathbf{v}\cdot\nabla_xp(t,\mathbf{x},\mathbf{v}) 
\nonumber \\
&-& \nabla_v\!\cdot[\mathbf{F}(C(t,\mathbf{x}))\!-\!k\mathbf{v}]\, p(t,\mathbf{x},\mathbf{v})\nonumber \\
&+& \frac{\sigma^2}{2}\Delta_v p(t,\mathbf{x},\mathbf{v}).\label{fpe}
\end{eqnarray}

\paragraph{Vessel extension, tip branching and anastomosis.} Tip branching and anastomosis contribute source and sink terms to the limiting equation for the tip density, as indicated in Appendix \ref{ap1}. The resulting equation is 
\begin{widetext}
\begin{eqnarray}
\frac{\partial}{\partial t} p(t,\mathbf{x},\mathbf{v})=
\frac{\alpha_1C(t,\mathbf{x})}{C_R+C(t,\mathbf{x})}\,
 p(t,\mathbf{x},\mathbf{v})\delta(\mathbf{v}-\mathbf{v}_0)
 - \gamma p(t,\mathbf{x},\mathbf{v}) \int_0^t \tilde{p}(s,\mathbf{x})\, ds  - \mathbf{v}\cdot \nabla_x   p(t,\mathbf{x},\mathbf{v}) \nonumber\\ + k \nabla_v \cdot (\mathbf{v} p(t,\mathbf{x},\mathbf{v})) - d_1\nabla_v \cdot
\left[\frac{\nabla C(t,\mathbf{x})}{[1+\gamma_1C(t,\mathbf{x})]^q}\, p(t,\mathbf{x},\mathbf{v})
\right]\! + \frac{\sigma^2}{2} \Delta_{v} p(t,\mathbf{x},\mathbf{v}),
\label{eq:final_strong_4}
\end{eqnarray}
\end{widetext}
where
\begin{equation}
\tilde{p}(t,\mathbf{x}) = \int p(t,\mathbf{x},\mathbf{v}')\,
d\mathbf{v}',\label{reduced_density}
\end{equation} 
is the marginal density  of $p(t,\mathbf{x},\mathbf{v})$.

Tip branching contributes the first term in the right hand side (RHS) of (\ref{eq:final_strong_4}). It is a birth term, $r_b(t)\, p$, with rate $r_b(t)$ proportional to the probability that a new branch be created at the interval $(t,t+dt)$ and to $\delta(\mathbf{v}-\mathbf{v}_0)$. The delta function recalls that new branches are created with velocity $\mathbf{v}_0$. Anastomosis occurs when a vessel tip meets a component of the vessel network that has been formed during previous times $0<s<t$. It contributes the second term in the RHS of (\ref{eq:final_strong_4}). It is a death term $r_d(t)\, p$, with rate proportional to the density of the vessel network, which is the integral of the marginal density up to time $t$. To further understand this, consider that the moving tip meets the past trajectory of a different tip at time $t$ in $(\mathbf{x},\mathbf{x}+d\mathbf{x})$. Let the time interval at which the other tip was in $(\mathbf{x},\mathbf{x}+d\mathbf{x})$ be $(s,s+ds)$. Clearly the destruction rate should be proportional to $\tilde{p}(s,\mathbf{x})\, ds$ provided we want to consider all possible tips with any velocities. Addition over all past times produces the overall death term. More formal mathematical arguments are given in Appendix \ref{ap1}. 

In appropriate limits, we may derive an integrodifferential equation for $\tilde{p}(t,\mathbf{x})$ from Eq. (\ref{eq:final_strong_4}). See e.g., \cite{BGr05,CBC09,alvaro13} for Chapman-Enskog derivations of similar balance equations describing nano devices. The balance equation for $\tilde{p}(t,\mathbf{x})$ will be nonlocal in time, thereby differing from balance equations for vessel densities postulated in the literature \cite{morale_chaplain_1998,morale:plank_sleeman:03,othmer,morale:chaplian_2006}.

\paragraph{Approximation of  the underlying field.} 
TAF diffuses and decreases where endothelial cells are present. Assuming that TAF consumption is only due to the new endothelial cells at the tips, the consumption is proportional to the velocity
$\mathbf{v}_i$ of the tip $i$ ($i=1,\ldots, N(t)$) in a region of infinitesimal radius about it. Then we have
\begin{eqnarray} \frac{\partial}{\partial t}C(t,\mathbf{x})\!&\!=\!& \! d_2 \Delta_x C(t,\mathbf{x})\nonumber \\ \!&\!-\!&\! \eta
C(t,\mathbf{x})\!\left|\frac{1}{N}\sum_{i=1}^{N(t)}
\mathbf{v}_i(t)\delta_N(\mathbf{x}-\mathbf{X}^i(t))\right|\!\!.\label{morale:eq_TAF}
\end{eqnarray}
Here $\delta_N(x)$ is a regularized smooth delta function (e.g., a Gaussian) that becomes $\delta(x)$ in the limit as $N\to\infty$. In this limit, the mean field term in this equation becomes the length of the tip flux and we obtain the following deterministic equation
\begin{eqnarray} \frac{\partial}{\partial t}C(t,\mathbf{x})
= d_2 \Delta_x C(t,\mathbf{x}) - \eta
C(t,\mathbf{x})|\mathbf{j}(t,\mathbf{x})| \label{morale:eq_TAF_lim},
\end{eqnarray}
where $\mathbf{j}(t,\mathbf{x})$ is the current density (flux)
vector at any point $\mathbf{x}$ and any time $t\geq 0$
\begin{equation}
\mathbf{j}(t,\mathbf{x})= \int \mathbf{v}'
p(t,\mathbf{x},\mathbf{v}')\, d \mathbf{v'}. \label{flux_max}
\end{equation}
The TAF production due to the tumor will be incorporated through a fixed flux boundary condition for (\ref{morale:eq_TAF_lim}).

\section{Boundary and initial conditions}
\label{sec:bc}
The system of equations (\ref{eq:final_strong_4}), (\ref{morale:eq_TAF_lim}) requires suitable initial and boundary conditions. We shall consider that angiogenesis occurs in two space dimensions.

Let $\mathbf{x}=(x,y)$ and $\mathbf{v}=(v,w)$. As said in the introduction, the tumor releases chemicals that attract blood vessels from a primary blood vessel towards it. A simple set up is to consider a two dimensional strip $\Omega=[0,L]\times\mathbb R \subset \mathbb R^2$ whose left boundary $\Omega_0=(0,y)$, $y\in\mathbb R$, is a  mature existing vessel (from which new vessels may sprout), whereas the right boundary $\Omega_L=(L,y)$, $y\in\mathbb R$, represents the tumor which is a source of  the TAF $C.$ Let $c_1(t,y)$ be the TAF flux emitted by the tumor at $x=L$. Appropriate boundary conditions for the underlying field $C$ that satisfies a parabolic equation are the Neumann conditions:
\begin{eqnarray}
&&\frac{\partial}{\partial n} C(t,0,y)=0, \quad
\frac{\partial}{\partial n} C(t,L,y)=\frac{c_1(y)}{d_2},\label{TAF_B_C}
\end{eqnarray}

 The boundary conditions for Equation (\ref{eq:final_strong_4}) should convey the idea that the vessel tips are issued at $x=0$, move and branch out more and more as $x$ changes from $x=0$ to $x=L$, and reach the tumor at the latter boundary. Except for the source term, Equation (\ref{eq:final_strong_4}) is a typical Fokker Planck parabolic equation having a noninvertible diffusion matrix: it has second order partial derivatives of $p$ with respect to the velocity but only first order partial derivatives with respect to position. Then we should impose
\begin{eqnarray}
p(t,\mathbf{x},\mathbf{v})\to 0 \mbox{ as } |\mathbf{v}|\to \infty,\label{bc_infty}
\end{eqnarray}
 but we cannot have proper Dirichlet or Neumann boundary conditions at $\Omega_0$ and $\Omega_L$ as Equation (\ref{eq:final_strong_4}) is only first order in the position coordinates. As it happens with the ``one-half boundary conditions'' for Boltzmann type equations (which are also first order in position),
 we should know $p$ at the boundaries $\Omega_0$ and $\Omega_L$ for vessel tips {\em entering} $\Omega$ ($\mathbf{v}\cdot\mathbf{n}<0$) in terms of $p$ for vessel tips {\em leaving} $\Omega$ ($\mathbf{v}\cdot\mathbf{n}>0$). Here $\mathbf{n}(\mathbf{x})$ is the unit vector normal to the boundary at a point $\mathbf{x}\in\partial\Omega$ and pointing outside the region $\Omega$. 

 To ascertain the proper boundary conditions at $\Omega_0$ and $\Omega_L$, we get a clue from problems of charge transport in semiconductor devices in which charge is injected at some boundaries and it is collected at others \cite{BGr05}. The key idea is that boundary conditions for $p$ having the above mentioned form should be compatible with physically meaningful conditions for appropriate moments of $p$ at the boundaries. In our case, it is reasonable to assume that we know the normal component of the flux (\ref{flux_max}) at the boundary $\Omega_0$ that emits tips and the marginal tip density at the tumor boundary $\Omega_L$: 
\begin{equation}
-\mathbf{n}\cdot\mathbf{j}(t;0,y)=  j_0(t,y),\quad \tilde{p}(t,L,y)=\tilde{p}_L(t,y), 
\label{flux_max_B_C}
\end{equation}
at  any time  $t \in [0, \infty).$ As $\mathbf{n}(\mathbf{x})$ is the unit vector normal to the boundary at a point $\mathbf{x}\in\partial\Omega$ and pointing outside the region $\Omega$, $\mathbf{n}\cdot\mathbf{j}>0$ (resp. $\mathbf{n}\cdot\mathbf{j}<0$) means that the flux is leaving (entering) $\Omega$. The normal flux {\em entering} the left boundary is given by the vessel production
\begin{eqnarray}\nonumber
&&-\frac{1}{N}\sum_{k=1}^\infty\int \mathbf{n}\!\cdot\!\mathbf{v} \frac{L}{|\mathbf{v}_0|}\, \alpha(C(t,x,y))\,\delta(\mathbf{x}-(0,y))\,\delta(\mathbf{v}-\mathbf{v}_0)\\
\nonumber 
&&\quad \times\,\delta(\mathbf{x}-\mathbf{X}^k(t))\,\delta(\mathbf{v}-\mathbf{v}^k(t))\, d\mathbf{x}\, d\mathbf{v},
\end{eqnarray} 
where $L$ is the distance to the tumor. In the mean field approximation, this expression becomes
\begin{equation}
j_0(t,y)=\frac{v_0\, L}{\sqrt{v_0^2+w_0^2}}\alpha(C(t,0,y))\, p(t,0,y,v_0,w_0), \label{J0}
\end{equation}
for a vector velocity $\mathbf{v}_0=(v_0,w_0)$.

As  far as  the boundary conditions on the density $p$, we assume that the density of vessel tips entering $\Omega$ is close to a local equilibrium distribution at the boundaries in such a way that the boundary conditions (\ref{flux_max_B_C}) are satisfied. Particular cases of such boundary conditions exist in the literature on Boltzmann type kinetic equations for semiconductors. Cercignani et al proposed charge neutrality and insulating boundary conditions for  the distribution function \cite{CGL01} (they credit a footnote in Baranger and Wilkins \cite{BW87} for the formulation of charge neutrality conditions). Bonilla and Grahn proposed injecting boundary conditions for a distribution function in \cite{BGr05}. The form of the  local equilibrium distribution may be postulated directly based on physical assumptions (as we do in this section) or obtained from an approximation of the distribution $p$ in some perturbative scheme \cite{CGL01,BGr05,CBC09,alvaro13}. To give simple examples of boundary conditions, let us assume that $p$ is close to a Maxwellian distribution with  temperature $\sigma^2/k$ and average velocity $\mathbf{v}_0$ at $\Omega_0$ and $\Omega_L$: 
\begin{eqnarray}
&& p^+(t,0,y,v,w)
=\frac{e^{-\frac{k|\mathbf{v}-\mathbf{v}_0|^2}{\sigma^2}}}{\int_0^{\infty}\!\int_{-\infty}^{\infty}
 v' e^{-\frac{k|\mathbf{v}'-\mathbf{v}_0|^2}{\sigma^2}}dv'\,dw'} \nonumber \\
&& \times\!  \left[j_0(t,y)\! -\! \int_{-\infty}^0\!\int_{-\infty}^{\infty}\!\!
v' p^-(t,0,y,v',w')d v' dw'\right]\!\!, \nonumber \\
&&p^-(t,L,y,v,w)=\frac{e^{-\frac{k|\mathbf{v}-\mathbf{v}_0|^2}{\sigma^2}}}{\int_{-\infty}^0\!\int_{-\infty}^{\infty}
e^{-\frac{k|\mathbf{v}'-\mathbf{v}_0|^2}{\sigma^2}}dv\,dw} \nonumber\\
&& \times  \!\left[\tilde{p}(t,L,y)\! -\!
\int_0^{\infty}\!\!\int_{-\infty}^{\infty}\! p^+(t,L,y,v',w')dv' dw'\right]\!\!. \label{p_B_C}
\end{eqnarray}
where $p^+=p$ for $v>0$ and $p^-=p$ for  $v<0$. The choice of boundary temperatures $\sigma^2/k$ corresponds to a dominant balance of the terms $k\nabla_v(\mathbf{v}p)$ and $\frac{1}{2}\sigma^2 \Delta_v p$ in (\ref{eq:final_strong_4}). Since all new vessels are assumed to branch with velocity $\mathbf{v}_0$, it is reasonable to assume that they also do so when they issue from the primary blood vessel at $x=0$. Thus we assume that the average velocity at $x=0$ is also $\mathbf{v}_0$.

We  may  notice  that (\ref{p_B_C}) implies
\begin{eqnarray}
&& \int_{0}^{\infty}\int_{-\infty}^{\infty} v p^+(t,0,y,v,w) dv\,
dw \nonumber \\
&&=\int_0^{\infty}\int_{-\infty}^{\infty}
\frac{v\,e^{-\frac{k|\mathbf{v}-\mathbf{v}_0|^2}{\sigma^2}}}{\int_0^{\infty}\int_{-\infty}^{\infty}dv'\,
dw'\, v'
e^{-\frac{k|\mathbf{v}'-\mathbf{v}_0|^2}{\sigma^2}}} \nonumber\\
&&\times \left[j_0(t,y) \right. \nonumber \\ && \left. \quad -\!
\int_{-\infty}^0\int_{-\infty}^{\infty} v' p^-(t,0,y,v',w')dv'
dw'\right]\! dv\, dw
\nonumber \\
&&=j_0(t,y)- \int_{-\infty}^0\int_{-\infty}^{\infty} v\,
p^-(t,0,y,v,w)\, dv\, dw,
 \label{p_B_C_1} \end{eqnarray}
 which is coherent with (\ref{flux_max}) and (\ref{flux_max_B_C}). 

Let us now assume that the two dimensional domain is a circular crown of radii $r_0<r<r_1$ centred at the origin. We  may  assume  that the outer boundary $|\mathbf{x}|=r_1$ describes a  mature existing vessel, from which new vessels may sprout, while the inner boundary $|\mathbf{x}|=r_0$ describes the tumor, i.e. a source of  the TAF $C.$  Boundary conditions for $C$ are similar to (\ref{TAF_B_C}) with radial derivatives at $r=r_1$ and $r=r_0$ as normal derivatives replacing those at $x=0$ and $x=L$, respectively. As in the case of the rectangular domain, we assume that we know the radial component of the current density vector {\em entering} the outer boundary,
\begin{eqnarray}
j_r(t,r_1,\theta)&=&j_1(t,\theta) \nonumber
\\ &=& v_{r0}\alpha(C(t,r_1,\theta))\, p(t,r_1,\theta,v_{r0},v_{\theta
0}), \label{J0rad}
\end{eqnarray}
for a vector velocity of radial and angular components $v_{r0}$ and $v_{\theta 0}$, respectively. The marginal density at the inner boundary (the tumor) $\tilde{p}(t,r_0,\theta)=\tilde{p}_0(t,\theta)$. Then the boundary conditions for $p$ are
\begin{eqnarray}
&&\!\!\!\!\!\!\!\!p^+(t,r_1,\theta,v_r,v_\theta)=\frac{e^{-\frac{k|\mathbf{v}-\mathbf{v}_0|^2}{\sigma^2}}}{\int_0^{\infty}\int_{-\frac{\pi}{2}}^{\frac{\pi}{2}}
 e^{-\frac{k|\mathbf{v}-\mathbf{v}_0|^2}{\sigma^2}}v_r^2 dv_r dv_\theta} \nonumber \\
 && \!\!\!\!\!\!\!\!\times\! \left[j_1(t,\theta)
 - \int_0^{\infty}\!\!\int_{\frac{\pi}{2}}^{\frac{3\pi}{2}} v_r^2 p^-(t,r_1,\theta,v_r,v_\theta)d
v_r dv_\theta\right]\!\! ,\nonumber\\
&&\!\!\!\!\!\!\!\!p^-(t,r_0,\theta,v_r,v_\theta)=\frac{e^{-\frac{k|\mathbf{v}-\mathbf{v}_0|^2}{\sigma^2}}}{\int_0^{\infty}\int_{\frac{\pi}{2}}^{\frac{3\pi}{2}}
v_r e^{-\frac{k|\mathbf{v}-\mathbf{v}_0|^2}{\sigma^2}}dv_r dv_\theta} \nonumber \\
&&\!\!\!\!\!\!\!\!\times\! \left[\tilde{p}_0(t,\theta)-
\int_0^{\infty}\!\!\int_{-\frac{\pi}{2}}^{\frac{\pi}{2}} p^+(t,r_0,\theta,v_r,v_\theta)v_r
dv_r dv_\theta\right]\!\!, \label{p_rad_B_C}
\end{eqnarray}
where $v_\theta$ is the angle formed by $\mathbf{v}$ with the inner radial direction pointing toward the origin, $p^+=p$ for $-\pi/2<v_\theta<\pi/2$ and $p^-=p$ for  $\pi/2<v_\theta<3\pi/2.$ Note that $\phi-\theta+v_\theta=\pi$ if the polar angles of the velocity and position vectors are $\phi$ and $\theta$, respectively.

\section{Numerical results}
\label{sec:numerical}
The parameter values we use when solving the model are given in Table \ref{table}. The values of $k$, $\sigma^2=k\tilde{v}_0^2$, $d_1C_R$ and $\tilde{v}_0=|\mathbf{v}_0|$ have been taken from Ref.~\cite{morale:Stokes_Lauffenburger:91}, $C_R$ is given in Ref.~\cite{morale:Stokes_Lauffenburger:91bis}. The tip birth rate $\alpha_1(t,\mathbf{x})$ is the probability per area per time that a new tip appears. Stokes and Lauffenburger estimated the probability per length per time from experiments on the inflammation-induced neo-vascularization of the rat cornea \cite{sholley}. They noted that 15 branches sprouted in 3 days from a 0.88 mm vessel \cite{sholley} and that half these branches could be assumed to be caused by branching and the other half by anastomosis. This gives a probability per length per time of $1.2\times 10^{-4}/\mu$m/hr \cite{morale:Stokes_Lauffenburger:91}.  Using Figures 1e and 1f in \cite{sholley}, we have counted 18 sprouts averaging 0.88 mm growth in 4 days and 11 sprouts averaging 0.54 mm growth in 4 days, respectively. The width of the cornea sector is about 1.9 mm which yields areas of 1.7 and 1 mm$^2$, respectively. Using Stokes and Lauffenburger's arguments, we find a probability per area per time of about $1.12\times 10^{-7}/\mu$m$^2$/hr in both cases. This is 31.1/m$^2$/s, the scale of $\alpha_1(t,\mathbf{x})$, which equals the coefficient $\alpha_1$ times the scale of $p(t,\mathbf{x},\mathbf{v})$. Using the value in Table \ref{table1}, we obtain $\alpha_1=1.538 \times 10^{-20}$ m$^2$/s$^3$. 
\begin{table}[ht]
\begin{center}\begin{tabular}{cccccccc}
 \hline
$\frac{1}{k}$& $\tilde{v}_0$ & $\sigma^2$&$\alpha_1$&$d_1C_R$ &$C_R$ &$\eta$ &$\gamma$\\
hr & $\frac{\mbox{$\mu$m}}{\mbox{hr}}$ & $10^{-21}\frac{\mbox{m}^2}{\mbox{s}^3}$& $10^{-20}\frac{\mbox{m}^2}{\mbox{s}^3}$ & $\frac{\mbox{$\mu$m$^2$}}{\mbox{hr}^2}$  & mol/m$^2$  & $\mu$m& $10^{-17}\frac{\mbox{m$^2$}}{\mbox{s$^2$}}$\\ 
8.5& 40 &4.035 & 1.538& 2400 & $10^{-16}$ & 4& 5.82\\
 \hline
\end{tabular}
\end{center}
\caption{Parameters used to solve the model equations. }
\label{table}
\end{table}

\begin{table}[ht]
\begin{center}\begin{tabular}{ccccccc}
 \hline
$\mathbf{x}$& $\mathbf{v}$ & $t$ &$C$& $p$ &$\tilde{p}$&$\mathbf{j}$\\
$L$ & $\tilde{v}_0$ & $\frac{L}{\tilde{v}_0}$ & $C_R$ & $\frac{1}{\tilde{v}_0^2L^2}$& $\frac{1}{L^2}$& $\frac{\tilde{v}_0}{L^2}$\\
mm&$\mu$m/hr & hr& mol/m$^2$&$10^{21}\frac{\mbox{s$^2$}}{\mbox{m$^4$}}$ & $10^{5}$m$^{-2}$&m$^{-1}$s$^{-1}$\\
$2$& 40 & 50 & $10^{-16}$ & 2.025 & 2.5 & 0.0028 \\ 
\hline
\end{tabular}
\end{center}
\caption{Units for nondimensionalizing the model equations. }
\label{table1}
\end{table}

We  have nondimensionalized the governing equations of our model, (\ref{eq:final_strong_4}) and (\ref{morale:eq_TAF_lim}), according to the units in Table \ref{table1}. The resulting nondimensional equations are
\begin{eqnarray} 
\frac{\partial p}{\partial t} &=& \frac{A\, C}{1+C}p\, \delta(\mathbf{v}-\mathbf{v}_0)
- \Gamma p\! \int_0^t \tilde{p}(s,\mathbf{x})\, ds  \nonumber\\
  &-& \mathbf{v}\cdot \nabla_x   p  
-\nabla_v \cdot\left[\!\left(\frac{\delta\, \nabla_x C}{(1+\Gamma_1C)^q}-\beta\mathbf{v}\right)\! p\right]  \nonumber \\
&+&  \frac{\beta}{2}\, \Delta_{v} p,\label{nondim_p}\\
\frac{\partial C}{\partial t} &=& \kappa\, \Delta_x C - \chi\,
C\, |\mathbf{j}| \label{nondim_C}.
\end{eqnarray}
The dimensionless parameters appearing in these equations are defined in Table \ref{table2}. $1/\kappa$ is the diffusive P\'eclet number, $\delta$ is the chemotactic responsiveness and $\beta$ is both a dimensionless friction coefficient and a noise diffusivity.

\begin{table}[ht]
\begin{center}\begin{tabular}{ccccccc}
 \hline
$\delta$ & $\beta$ &$A$& $\Gamma$& $\Gamma_1$ &$\kappa$&$\chi$\\
 $\frac{d_1C_R}{\tilde{v}_0^2}$ & $\frac{kL}{\tilde{v}_0}$ & $\frac{\alpha_1L}{\tilde{v}_0^3}$ & $\frac{\gamma}{\tilde{v}_0^2}$&$\gamma_1C_R$& $\frac{d_2}{\tilde{v}_0 L}$& $\frac{\eta}{L}$\\
1.5 & 5.88 & $22.42$ & 0.3 & 1& $0.0045$ & 0.002 \\
 \hline
\end{tabular}
\end{center}
\caption{Dimensionless parameters. } 
\label{table2}
\end{table}

We now write the boundary conditions in nondimensional form for the strip geometry $0<x<1$, $y\in\mathbb{R}$. The boundary conditions for $C$ are
\begin{eqnarray}
\frac{\partial C}{\partial x}(t,0,y)=0, \quad\frac{\partial C}{\partial x}(t,1,y)=f(y), \, \lim_{y\to\pm\infty}C= 0, \label{nondim_bc_C}
\end{eqnarray}
where $f(y)=L\, c_1(Ly)/(C_Rd_2)$ is a nondimensional flux. We have used $c_1(y)=a\, e^{-y^2/b^2}$, with $a=5.5\times 10^{-27}$ mol/(m$^2$s), $d_2=10^{-13}$ m$^2$/s, and $b=0.4$ mm ($b$ is about half the assumed tumor size). The initial condition for the TAF concentration is the Gaussian 
\begin{eqnarray}
C(0,x,y)=1.1\, C_R e^{-[(x-L)^2/c^2+y^2/b^2]},\label{icC}
\end{eqnarray} 
with $c=3$ mm, whereas the initial vessel density is 
\begin{eqnarray}
p(0,x,y,v,w)= \frac{2N_0}{\pi^2l_xl_y\tilde{v}_0^2}\,e^{-x^2/l_x^2-y^2/l_y^2-|\mathbf{v}-\mathbf{v}_0|^2/\tilde{v}_0^2},\label{icP}
\end{eqnarray} 
$l_y=10 l_x=0.8$ mm, that corresponds to $N_0=20$ initial vessel tips. Nondimensionalization of the initial conditions (\ref{icC}) and (\ref{icP}) by using Table \ref{table1} is obvious. In nondimensional form, the boundary conditions (\ref{p_B_C}) for $p$ are
\begin{eqnarray}
&& p^+(t,0,y,v,w)
=\frac{e^{-|\mathbf{v}-\mathbf{v}_0|^2}}{\int_0^{\infty}\!\int_{-\infty}^{\infty}
 v' e^{-|\mathbf{v}'-\mathbf{v}_0|^2} dv'\,dw'} \nonumber \\
&& \times\!  \left[j_0(t,y)\! -\! \int_{-\infty}^0\!\int_{-\infty}^{\infty}\!\!
v' p^-(t,0,y,v',w')d v' dw'\right]\!  \label{nondim_bc_p0}
 \end{eqnarray}
for $x=0$ and $v>0,$
\begin{eqnarray}
&&p^-(t,1,y,v,w)=\frac{e^{-|\mathbf{v}-\mathbf{v}_0|^2}}{\int_{-\infty}^0\!\int_{-\infty}^{\infty}
e^{-|\mathbf{v}'-\mathbf{v}_0|^2}dv'\,dw'} \nonumber\\
&& \times  \!\left[\tilde{p}(t,1,y)\! -\!
\int_0^{\infty}\!\!\int_{-\infty}^{\infty}\! p^+(t,1,y,v',w')dv' dw'\right]\!
\label{nondim_bc_pL}
\end{eqnarray}
for $x=1$ and $v<0$. Eq.\! (\ref{J0}) produces the nondimensional flux $j_0$:
\begin{eqnarray}
j_0(t,y)=A\, v_0\frac{C}{1+C}\, p(t,0,y,v_0,w_0)   \label{nondim_j0}
\end{eqnarray}
($\sqrt{v_0^2+w_0^2}=|\mathbf{v}_0|^2=1$ in nondimensional units).

We have solved (\ref{nondim_p})-(\ref{nondim_bc_pL}) by an explicit finite-difference scheme, using upwind differences for positive $v$ and $w$ and downwind differences for negative $v$ and $w$. The boundary conditions (\ref{nondim_bc_p0}) and (\ref{nondim_bc_pL}) then give the needed boundary value of $p^\pm$ at one time step in terms of the value of $p^\mp$, which is known at the precedent time step. The integrals are approximated by the composite Simpson rule and $\delta(\mathbf{v}-\mathbf{v}_0)$ in (\ref{nondim_p}) is approximated by a Gaussian. 

\begin{figure}[ht]
\begin{center}
\includegraphics[width=9cm]{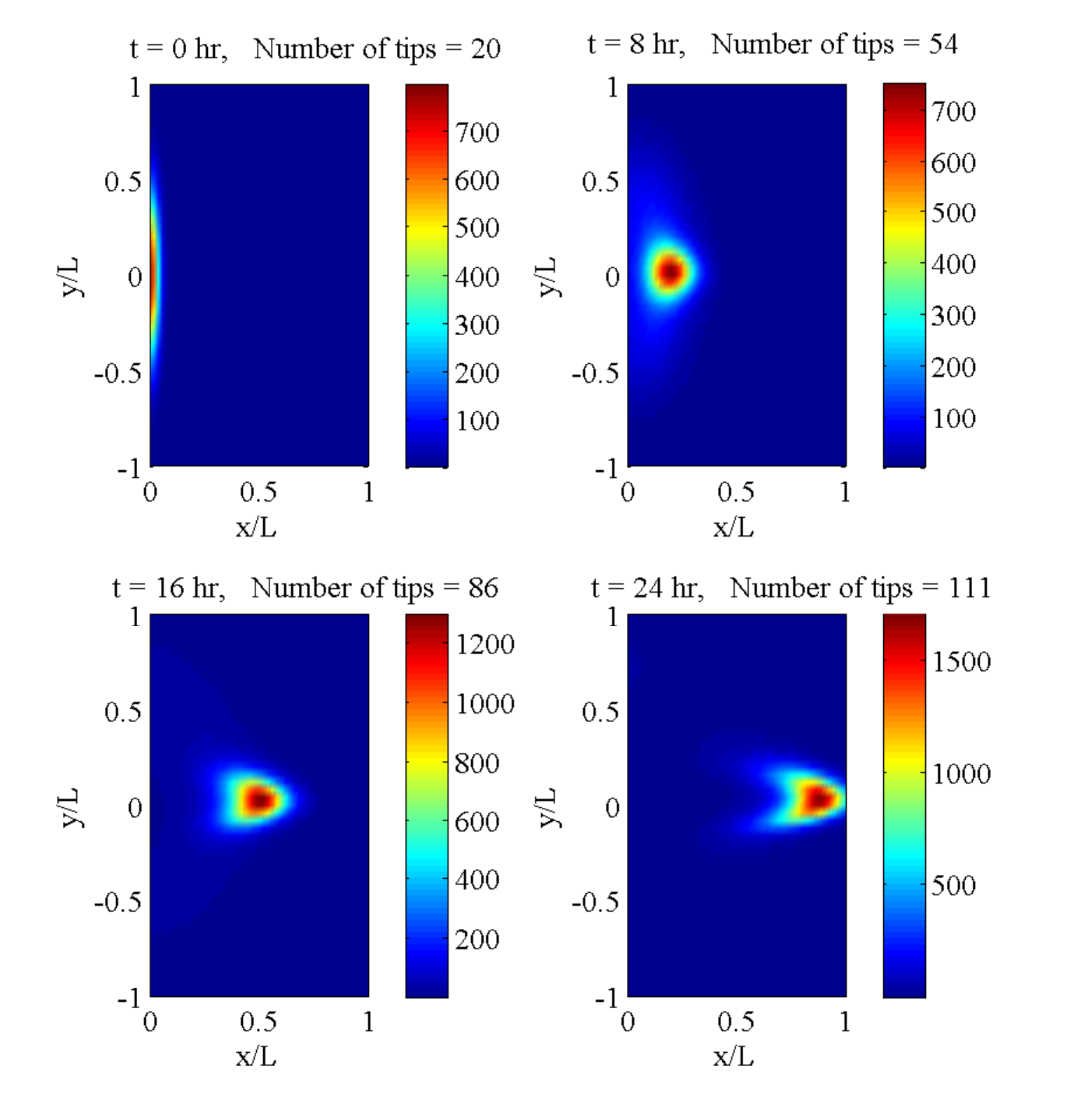} \qquad
\end{center}
\caption{Density plot of the marginal tip density $\tilde{p}(t,x,y)$ for different times showing how tips are created and march toward the tumor. Nondimensional parameter values are as in Table \ref{table2}.
\label{tip_density_densityplot}}
\end{figure}
\begin{figure}[ht]
\begin{center}
\includegraphics[width=9cm]{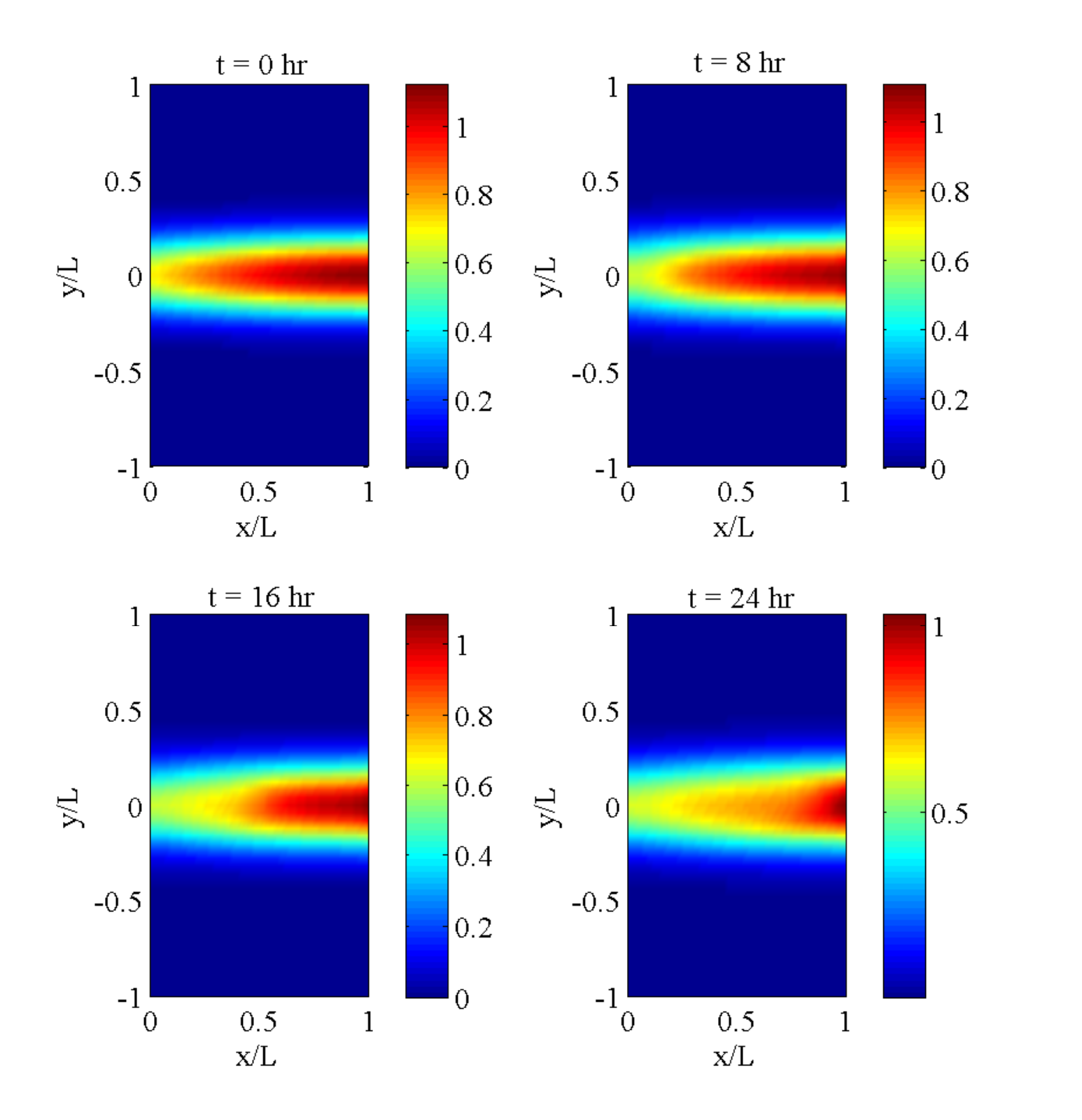} \qquad
\end{center}
\caption{Density plot of the TAF concentration $C(t,x,y)/C_R$ for different times showing how tips consume TAF in their march towards the tumor. Nondimensional parameter values are as in Table \ref{table2}.
\label{TAF_densityplot}}
\end{figure}

\begin{figure}[ht]
\begin{center}
\includegraphics[width=9cm]{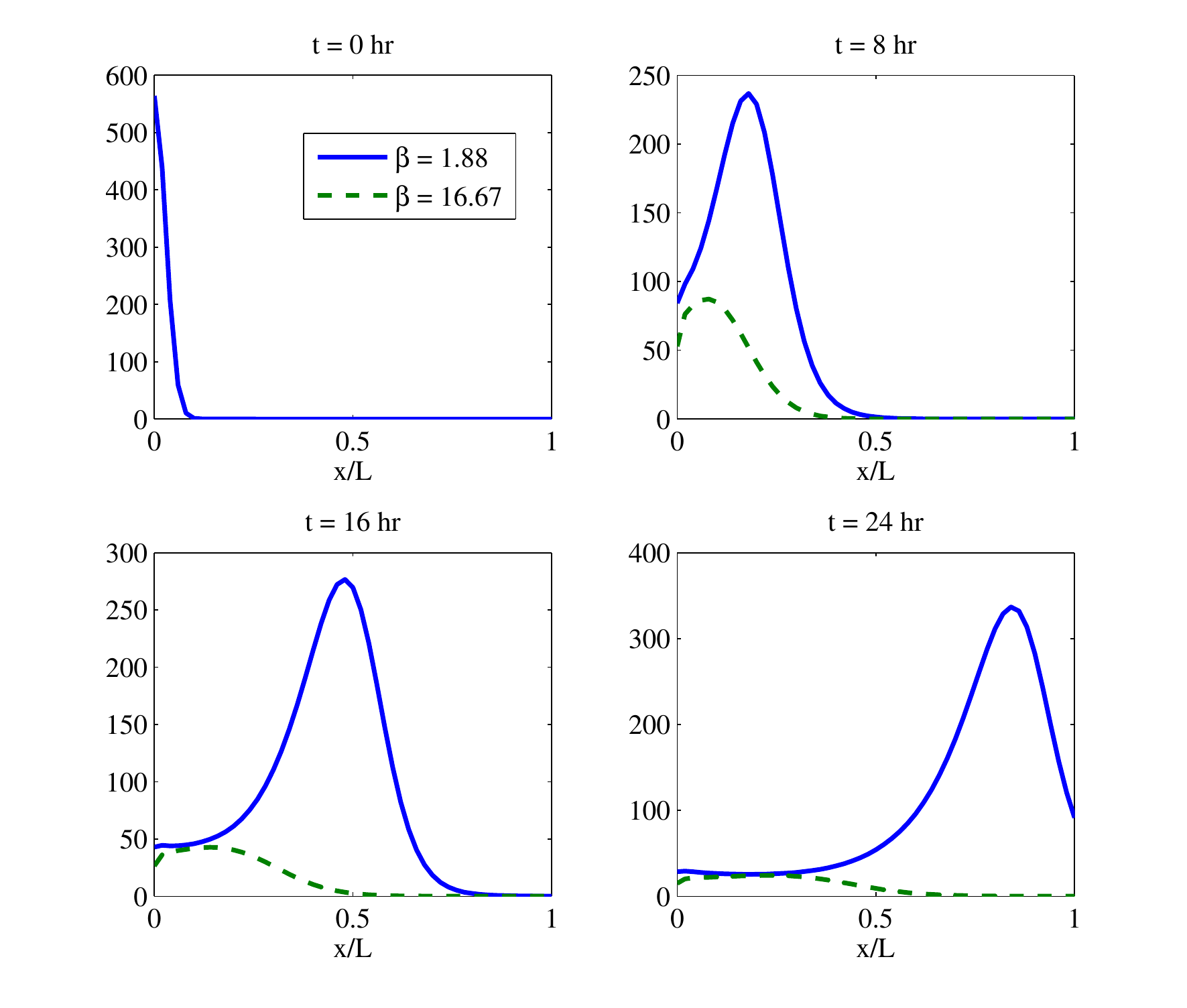} \qquad
\end{center}
\caption{The integrated marginal tip density profile $\int_{-\infty}^\infty\tilde{p}(t,x,y)\, dy$ for different times showing that the tip generation and motion proceeds as a pulse that grows as it advances towards the tumor. Persistence times $1/k$ are 8.5 hours (solid line, $\beta=5.88$) and 3 hours (dashed line, $\beta=16.67$). Larger $\beta$ values result in arresting the motion of the vessel tips toward the tumor.}   
\label{tip_wave}
\end{figure}

The numerical solution of (\ref{nondim_p})-(\ref{nondim_bc_pL}) depicted in Figure \ref{tip_density_densityplot} shows that vessel tips are created at $x=0$ and move towards the tumor at $x=1$ ($L$ is 2 mm in dimensional units). The total tip number, $N(t)$, is the integer part of the mass, $\int\tilde{p}(t,\mathbf{x})\, d\mathbf{x}$ and it increases with time. As shown in Figure \ref{TAF_densityplot}, the vessel tips consume TAF as they move. Figure \ref{tip_wave} indicates that the marginal tip density $\int_{-\infty}^\infty\tilde{p}(t,x,y)dy$ advances as a growing pulse wave. At each fixed $x>0$, the tip density is very small before new tips arrive from the left. Then TAF is consumed, new tips are created and this density increases. No new tips are created after TAF disappears but the sink term in the right side of (\ref{nondim_p}) continues tip destruction: $p$ decays and the pulse has then passed the vertical line at $x$.

\begin{figure}[ht]
\begin{center}
\includegraphics[width=9cm]{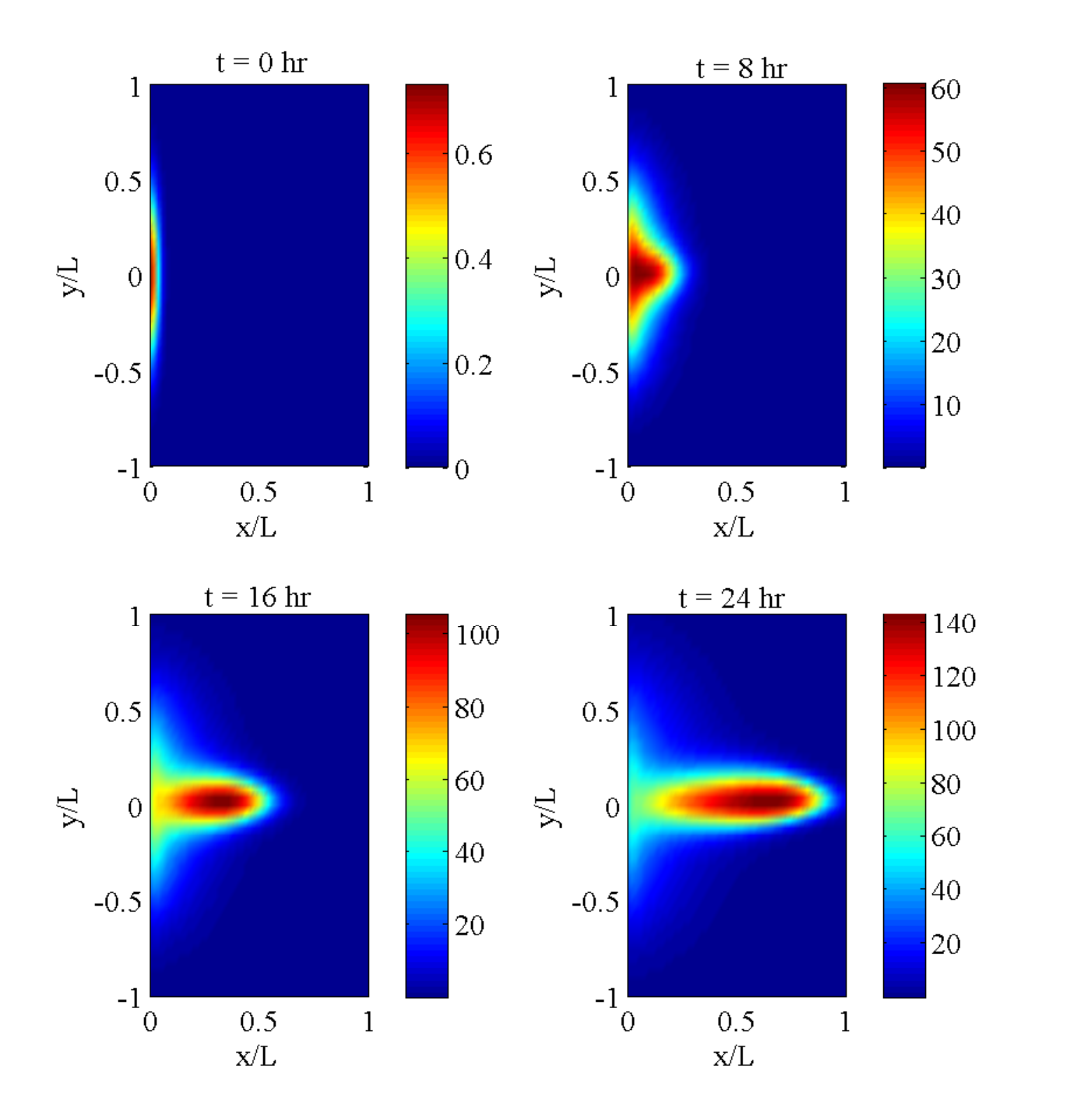} \qquad
\end{center}
\caption{Density plot of the vessel network left by the tips, $\int_0^t\tilde{p}(s,x,y)\, ds$, for different times. Nondimensional parameter values are as in Table \ref{table2}.
\label{vessel_densityplot}}
\end{figure}

The density plot of $\int_0^t\tilde{p}(s,x,y)\, ds$ (vessel network density) in Figure \ref{vessel_densityplot}, shows how the created tips form a growing vessel network that moves towards the tumor. 
The behavior of the angiogenic vessel network depends very much on the values of the dimensionless parameters in Table \ref{table2}. These parameters should ideally be fit from experiments and, in this respect, a series of vessel images taken several times a day would be most helpful. From measurements in \cite{morale:Stokes_Lauffenburger:91bis,morale:Stokes_Lauffenburger:91}, the persistence time $1/k$ and the velocity $\tilde{v}_0$ (and therefore $\beta$) vary appreciably depending on conditions met by endothelial cells. The friction force $-\beta\mathbf{v}$ opposes the chemotactic force $\delta\nabla_xC/(1+\Gamma_1C)^q$ that drives the vessel network towards the tumor. For large values of $\beta$ (small values of $\nabla_xC$), the vessel tips stop and may even move back, so that they never arrive at the tumor. Anti-angiogenic therapies may target increasing $\beta$ or decreasing the chemotactic force (decreasing $\nabla_xC$ may be achieved by increasing $c$ in (\ref{icC})). Pro-angiogenic therapies may have the opposite targets. In Fig. \ref{tip_wave}, we have also depicted the arresting effect that increasing $\beta$ has on the vessel network. In experiments, different drugs have the effect of arresting the vessel network before it arrives at the target area and thinning it, as shown in Figure \ref{angio_real2}. We also observe that the treatments inhibit vessel proliferation near the primary vessel. This effect might be achieved by tuning the parameter $A$ that controls vessel tip production both in (\ref{nondim_p}) and in the boundary condition (\ref{nondim_bc_p0}) through (\ref{nondim_j0}). Smaller $A$ results in less production of vessels. The parameters $\Gamma$ and $\chi$ have opposite effects to those of $A$. 

A possible program to use our model to test anti- or pro-angiogenic substances could consist of the following. Firstly calibrate the model by a number of experiments. Secondly, ascertain whether drugs can be used to tune parameters of the model and to attain anti- or pro-angiogenic effects. Finally solve numerically the model equations, obtain and test predictions thereof by measuring TAF concentration and marginal vessel density. The latter could be ascertained from images of the network at successive times such as those in Figure \ref{angio_real2}. Of course, there are several simplistic features in the model that may need to be reconsidered. Obvious ones are that there is an additional haptotactic force driving vessels toward the tumor \cite{VK_morale_jomb}. Blood perfusion in newly formed vessels needs to be considered and the effect of vessel retraction due to low blood circulation included in the model. This latter issue could be included along the lines of Ref. \cite{morale:chaplian_2006}.

\section{Conclusions}
\label{sec:conclusions}
We have derived equations for the density of vessel tips and for the TAF density during tumor-driven angiogenesis on the basis of a hybrid model. In this model, the tips undergo a stochastic process of tip branching, vessel extension and anastomosis whereas TAF is described by a reaction-diffusion equation with a sink term proportional to the average tip flow. In a limit of sufficiently many tips, the tip density satisfies a Fokker-Planck type equation coupled to a reaction-diffusion equation for the TAF density. We have proposed boundary conditions for these equations which describe the flux of vessel tip injected from a primary blood vessel in response to TAF emitted by the tumor and the tip density eventually arriving at the tumor. Numerical solution of the model in a simple geometry shows how tips are created at the primary blood vessel, propagate and proliferate towards the tumor and may or not reach it after a certain time depending on the parameter values. This is consistent with the known biological facts and with the original stochastic equations. 

Additional work exploring the relation between our model and the stochastic equations is left for the future. Although the mean field continuum model should describe well average behavior, we expect that the stochastic description (from which the continuum model is derived) presents large variance in regions where the number of tips is small. The stochastic density of vessels has a large variance close to the initiating primary blood vessel, whereas fluctuations become unimportant closer to the tumor, in a region with many more vessels. Then the stochastic density of vessels derived from the solution of the fully stochastic model will approach the mean vessel density studied in this paper and represented in Fig. \ref{vessel_densityplot}. The evolution of the vessel network depends on the values of the parameters in the model and a thorough study is required to design strategies based on modifying them. Ultimately, the effects of haptotaxis (fibronectin, MDE) and blood perfusion in the vessels may have to be added to the model in order to improve it.

\acknowledgments
This work has been supported by the Spanish Ministerio de Econom\'\i a y Competitividad grant
FIS2011-28838-C02-01. VC  has been supported  by a Chair of Excellence at the Universidad Carlos III de Madrid. It is a great pleasure to acknowledge fruitful discussions with Elisabetta Dejana
of the Institute FIRC of Molecular Oncology  of  Milan, and Daniela Morale of the Department of Mathematics of the University of Milan. We also thank Elisabetta Dejana for allowing us to use Figure \ref{angio_real2}.

\appendix
\section{Derivation of the equation for the tip density}
\label{ap1}
We need to introduce some notation. The union of the trajectories of the $N(t)$ tips that exist up to time $t$,
\begin{equation}\mathbf{X}(t)=\displaystyle\bigcup_{i=1}^{N(t)}
\{\mathbf{X}^i(s), T^i\le s\le \min\{t, \Theta_i \} \}
\label{network}
\end{equation} 
is the network of endothelial cells. Here $T^i$ and $\Theta"$ are the random birth (by branching) and death (by anastomosis) times of the $i$th tip. Each particle tip is characterized by its space $\mathbf{X}^k(t)$ and velocity $\mathbf{v}^k(t)$ coordinates, so that the whole process is characterized by the stochastic processes
$\{(\mathbf{X}^k(t),\mathbf{v}^k(t)),\, k = 1, . . . ,N(t),\, t\in\mathbb R_+\}$. 

At any time $t\geq 0$, the number of tips, $N(t)$, is of the same order  $O(N)$, where $N$ is a large positive integer. There are two fundamental random spatial measures describing the system at time $t$. Let $Q_N(t)(A)$ be the number of tips with positions and velocities in the phase space region $A$ at time $t$ divided by $N$. Formally, the empirical measure $Q_N$ of the processes $(\mathbf{X}^k(t),\mathbf{v}^k(t)),\, k=1,\ldots, N(t)$ is defined as
\begin{equation}
Q_N(t):=\frac{1}{N}
\sum_{k=1}^{N(t)}\epsilon_{(\mathbf{X}^k(t),\mathbf{v}^k(t))}.
\label{def: _empirical measure_global}
\end{equation}
Here $\epsilon_{(\mathbf{X}^k(t),\mathbf{v}^k(t))}(A)=\int_A \delta(\mathbf{x}-\mathbf{X}^k(t))\,\delta(\mathbf{v}-\mathbf{v}^k(t))\,d\mathbf{x}d\mathbf{v}$ and the delta function is the generalized derivative of the Dirac measure $\epsilon_{(\mathbf{X}^k(t),\mathbf{v}^k(t))}$. If we count tips that are in a spatial region at time $t$, no matter their velocities, their random empirical distribution $T_N(t)$ is given by
\begin{equation}
T_N(t)=\frac{1}{N} \sum_{k=1}^{N(t)}\epsilon_{X^k(t)}= Q_N(t)(\cdot\times\mathbb R^d). 
\label{def: _empiricalmeasure_tips}
\end{equation}
Under appropriate conditions, we have 
\begin{eqnarray}
&&Q_N(t)(d(\mathbf{x},\mathbf{v}))\sim
p(t,\mathbf{x},\mathbf{v})\, d\mathbf{x}d\mathbf{v},\label{def_density_p}\\
&&T_N(t)(d(\mathbf{x},\mathbf{v}))\sim\tilde{p}(t,\mathbf{x})\, d\mathbf{x}.\label{def_density_ptilde}
\end{eqnarray}

\subsubsection{Vessel extension}
Using the empirical measure $Q_N(t)$ of (\ref{def: _empirical measure_global}), we can write (\ref{gN}) as
\begin{eqnarray}
&&\int\! g(\mathbf{x},\mathbf{v})\, Q_N(t)(d(\mathbf{x},\mathbf{v}))=\!\int\! g(\mathbf{x},\mathbf{v})\, Q_N(0)(d(\mathbf{x},\mathbf{v}))\nonumber \\
&&
+ \int_0^t\!\!\int\!\frac{1}{N} \sum_{k=1}^{N(s)}\!\mathbf{v}^k(s)
\cdot\nabla_xg(\mathbf{X}^k(s),\mathbf{v}^k(s))\,ds \nonumber \\
 &&+ \int_0^t\!\frac{1}{N} \sum_{k=1}^{N(s)} [\mathbf{F}(C(\mathbf{X}^k(s)))-k\mathbf{v}^k(s)] \nonumber \\
  && \quad \quad \cdot\nabla_v g(\mathbf{X}^k(s),\mathbf{v}^k(s))\,ds
 \nonumber\\
&&\!+\frac{\sigma^2}{2}\! \int_0^t\!\!\frac{1}{N} \sum_{k=1}^{N(s)}
\Delta_v g(\mathbf{X}^k(s),\mathbf{v}^k(s))\, ds +\tilde{M}_{1,N}(t).
\label{gQN}
\end{eqnarray}

\subsubsection{Addition of tip branching}
Let us denote by $\Phi(ds\times d\mathbf{x}\times d\mathbf{v})$ the random variable that counts those tips born from an existing tip during times on $(s,s+ds]$, with positions on $(\mathbf{x},\mathbf{x}+d\mathbf{x}]$, and velocities on $(\mathbf{v},\mathbf{v}+d\mathbf{v}]$. Tip branching, described by the scaled marked point process $\Phi_N=N^{-1}\Phi$, contributes an additional term to (\ref{gQN}):
\begin{eqnarray}
&& \int_0^t\!\int\!g(\mathbf{x},\mathbf{v})\Phi_N(ds\times d\mathbf{x}\times d\mathbf{v}) \delta(\mathbf{v}-\mathbf{v}_0) \nonumber\\
&& =\int_0^t\!\int\!g(\mathbf{x},\mathbf{v})
\alpha(C(s,\mathbf{x}))\delta(\mathbf{v}-\mathbf{v}_0)
 Q_N(s)(d(\mathbf{x},\mathbf{v}))\, ds \nonumber\\
 && \quad \quad +\tilde{M}_{2,N}(t)\label{g2}
\end{eqnarray}
(see  e.g.\! \cite{morale:B}, p.235), where
\begin{eqnarray}
&&\tilde{M}_{2,N}(t)= \!\int_0^t\!\! \int g(\mathbf{x},\mathbf{v})
[\Phi_N(ds\times d\mathbf{x}\times d\mathbf{v}) \delta(\mathbf{v}-\mathbf{v}_0) \nonumber\\
&&\quad\quad\quad- \alpha(C(s,\mathbf{x}))\delta(\mathbf{v}-\mathbf{v}_0)
 Q_N(s)(d(\mathbf{x},\mathbf{v}))ds],\label{mtilde_2}
\end{eqnarray}
is a zero mean martingale.

\subsubsection{Addition of anastomosis}
Let us denote by $\Psi(ds\times d\mathbf{x}\times d\mathbf{v})$ the random variable that counts those tips which are absorbed by the existing vessel network during time $(s,s+ds],$ with  position in
$(\mathbf{x}, \mathbf{x}+d\mathbf{x}],$ and velocity in $(\mathbf{v}, \mathbf{v}+d\mathbf{v}].$ The contribution from the death process described by the scaled marked point process $\Psi_N:=N^{-1} \Psi$ is (see e.g.\ \cite{morale:B}, p.235 or \cite{karlin2}, p.252)
\begin{eqnarray}
&& \int_0^t\!\int\!g(\mathbf{x},\mathbf{v})\Psi_N(ds\times d\mathbf{x} \times d\mathbf{v}) \nonumber\\
&=& \int_0^t\!\int\!g(\mathbf{x},\mathbf{v}) \frac{\gamma}{N}
\delta(\mathbf{x}-\mathbf{X}(s))Q_N(s)(d(\mathbf{x},\mathbf{v}))ds
\nonumber \\
&+& \tilde{M}_{3,N}(t),\label{g3}
\end{eqnarray}
where $\delta(\mathbf{x}-\mathbf{X}(t))$ is given by 
\begin{eqnarray}\label{history}
\delta(\mathbf{x}-\mathbf{X}(t))&=& \displaystyle \int_0^t
ds\sum_{i=1}^{N(s)} \,\delta(\mathbf{x}-\mathbf{X}^i(s)), 
\end{eqnarray}
 and 
\begin{eqnarray}
\tilde{M}_{3,N}(t)&\!=\!&\!\int_0^t\!\!  \int g(\mathbf{x},\mathbf{v})
[\Psi_N(ds\times d\mathbf{x} \times d\mathbf{v})\nonumber\\
&\!-\!&\! \left.\frac{\gamma}{N}
\delta(\mathbf{x}-\mathbf{X}(s))Q_N(s)(d(\mathbf{x},\mathbf{v}))ds\right]\!\!,\label{mtilde_3}
\end{eqnarray}
is itself a zero mean martingale. The delta function (\ref{history}) indicates whether a tip has passed through the point $\mathbf{x}$ during any time up to $t>0$. This can be formally justified as follows. The Hausdorff measure associated with the stochastic network $\mathbf{X}(t)$ of (\ref{network}) can be expressed in terms of the occupation time of a spatial region (a planar Borel set) by tips that exist up to a time $t>0$ (see page 225 of \cite{protter} or page 252 of \cite{karlin2} for  the particular case of  SDE's  driven by the classical Brownian  motion):
\begin{eqnarray}
\mathcal H^1 (\mathbf{X}(t)\cap A) &=& \displaystyle \int_0^t ds\sum_{i=1}^{N(s)}  \, \mathbb I_A(\mathbf{X}^i(s))  \nonumber \\
&=&\displaystyle \int_0^t
ds\sum_{i=1}^{N(s)}\epsilon_{\mathbf{X}^i(s)} (A),
\label{occupation_time}
\end{eqnarray}
where $\mathbb I_A(\mathbf{x})=1$ if $\mathbf{x}\in A$, 0 otherwise. As the tip trajectories are sufficiently regular due to the choice (\ref{eq:langevin}) of a Langevin model for the vessels extensions, the generalized derivative of the measure (\ref{occupation_time}) is (\ref{history}), as introduced in \cite{capasso_villa_2008}. In practice, the delta functions in Equations (\ref{g3})-(\ref{mtilde_3}) are regularized (e.g., they are Gaussian functions) and become delta functions only in the limit as $N\to\infty$. 

Summing up (\ref{g}),(\ref{g2}), (\ref{g3})  and (\ref{mtilde}) with
(\ref{mtilde_2}), (\ref{mtilde_3}) we get
\begin{widetext}
\begin{eqnarray}
&&\int\! g(\mathbf{x},\mathbf{v})\, Q_N(t)(d(\mathbf{x},\mathbf{v}))
=  \!\int\! g(\mathbf{x},\mathbf{v})\,
Q_N(0)(d(\mathbf{x},\mathbf{v})) +\int_0^t\!\int\!\mathbf{v}\cdot\nabla_xg(\mathbf{x},\mathbf{v})
Q_N(s)(d(\mathbf{x},\mathbf{v}))\, ds \nonumber\\
&&\, +\!\! \int_0^t\!\!\!\int\!
[\mathbf{F}(C(s,\mathbf{x}))-k\mathbf{v}]\!\cdot\!\nabla_v
g(\mathbf{x},\mathbf{v}) Q_N(s)(d(\mathbf{x},\mathbf{v}))\, ds
+ \int_0^t\int\! \frac{\sigma^2}{2}\Delta_v g(\mathbf{x},\mathbf{v})
Q_N(s)(d(\mathbf{x},\mathbf{v})) ds\nonumber\\
&&\, + \int_0^t \! \!\int\!\alpha(C(s,\mathbf{x}))\delta(\mathbf{v}-\mathbf{v}_0)Q_N(s)(d(\mathbf{x},\mathbf{v}))\, ds - \int_0^t\!\! \int\! \frac{\gamma}{N} \delta(\mathbf{x}-\mathbf{X}(s))
g(\mathbf{x},\mathbf{v}) Q_N(s)(d(\mathbf{x},\mathbf{v}))\,
ds + \tilde{M}_N(t),\label{g_sum_bis}
\end{eqnarray}
\end{widetext}

where  now
$$\tilde{M}_N(t)=  \tilde{M}_{1,N}(t) +    \tilde{M}_{2,N}(t)   +    \tilde{M}_{3,N}(t)$$
is still a zero mean martingale.

By suitable laws of large  numbers, whenever $N$ is sufficiently large, $Q_N$ may admit a density given by (\ref{def_density_p}) \cite{oel2,sznitman}. Consequently, $\delta(\mathbf{x}-\mathbf{X}(t))$ in (\ref{history}) approaches its mean value \cite{AKV_2009}
\begin{eqnarray}
\lambda(t,\mathbf{x})= \displaystyle \int_0^t \tilde{p}(s,\mathbf{x})\, ds,
\end{eqnarray}
where $\tilde{p}$ is the marginal tip density (\ref{reduced_density}). We now integrate by parts (\ref{g_sum_bis}), differentiate the result with respect to time and ignore the martingales in the limit as $N\to \infty$, thereby obtaining (\ref{eq:final_strong_4}). 

A rigorous derivation of (\ref{eq:final_strong_4}) from (\ref{g_sum_bis}) requires additional mathematical analysis including a proof of existence, uniqueness, and sufficient regularity  of the  solution  of (\ref{eq:final_strong_4}) subject to suitable boundary and initial conditions; see  also \cite{oel2,B_C_M}. This is outside the scope of the present paper.

\end{document}